# Spin Seebeck effect from antiferromagnetic magnons and critical spin fluctuations in epitaxial $FeF_2$ films

Junxue Li[1], Zhong Shi[2], Victor H. Ortiz[1], Mohammed Aldosary[1,3], Cliff Chen[1], Vivek Aji[1], Peng Wei[1], and Jing Shi[1]

1. Department of Physics and Astronomy, University of California, Riverside, California 92521, USA
2. School of Physics Science and Engineering, Tongji University, Shanghai 200092, China
3. Department of Physics and Astronomy, King Saud University, Riyadh 11451, Saudi Arabia

We report a longitudinal spin Seebeck effect (SSE) study in epitaxially grown $FeF_2$(110) antiferromagnetic (AFM) thin films with strong uniaxial anisotropy over the temperature range of 3.8 - 250 K. Both the magnetic field- and temperature-dependent SSE signals below the Néel temperature ($T_N$=70 K) of the $FeF_2$ films are consistent with a theoretical model based on the excitations of AFM magnons without any net induced static magnetic moment. In addition to the characteristic low-temperature SSE peak associated with the AFM magnons, there is another SSE peak at $T_N$ which extends well into the paramagnetic phase. All the SSE data taken at different magnetic fields up to 12 T near and above the critical point $T_N$ follow the critical scaling law very well with the critical exponents for magnetic susceptibility of 3D Ising systems, which suggests that the AFM spin correlation is responsible for the observed SSE near $T_N$.

Antiferromagnetic insulators (AFMI) have recently attracted a great deal of interest in the emerging field of antiferromagnetic spintronics due to their unique properties such as robustness against magnetic field perturbation and ultrafast spin-dynamics [1-4]. In early reports, thin AFMI layers are found to enhance spin current transmission when they are inserted between a ferrimagnetic insulator (FMI) and a heavy metal (HM), such as the NiO and CoO AFMI layers in $Y_3Fe_5O_{12}$(YIG)/NiO/Pt [5, 6] and YIG/CoO/Pt [6], where an increased spin Seebeck effect (SSE) signal is attributed to the enhanced spin conductance in the AFMI spacer around its phase transition temperature [6, 7]. In addition, AFMI themselves are reported as sources of pure spin current under an applied magnetic field which generate SSE signals in AFMI/HM heterostructures such as $MnF_2$/Pt [8], $Cr_2O_3$/Pt [9], and $\gamma$-$Fe_2O_3$/Pt [10].

The origin of SSE in AFMI-based heterostructures has been under active investigations. On one hand, the SSE signal in $Cr_2O_3$/Pt under a strong magnetic field is found to be proportional to the net equilibrium magnetization of $Cr_2O_3$ [9], i.e., it is negligibly small until the magnetic field exceeds the spin-flop transition field ($H_{SF}$) and produces a net induced magnetic moment. On the other hand, a finite SSE signal is reported in $MnF_2$/Pt when the magnetic field is less than $H_{SF}$, indicating that the field-induced magnetic moment is not the only cause of the SSE in AFMI [8]. In ferromagnetic insulator/HM heterostructures, SSE below the Curie temperature is usually attributed to the spin current due to magnon flow driven by a temperature gradient [11-15]. In a uniaxial antiferromagnet, there are two branches of magnons, namely $\alpha$- and $\beta$-mode magnons (Fig. 1(a)), which have opposite chiralities carrying opposite angular momenta [16-19]. At zero magnetic field, both modes are degenerate; therefore, there is no net magnon flow until a magnetic field is applied along the AFM spin direction [18, 20, 21], as shown in Fig. 1(b). In a recent theoretical model proposed by Rezende *et al.* [20, 21], a field-induced AFM magnon imbalance can lead to a characteristic SSE peak in the absence of any net equilibrium magnetization (Fig. 1(c)). This AFM magnon picture explains the low-field SSE signal in $MnF_2$/Pt. Clearly, a unified picture is lacking regarding the low-field SSE responses. In order to clarify the physical origin of the SSE in AFMI, a uniaxial AFMI with an unusually high spin-flop field is desired so that no equilibrium magnetization is induced with any laboratory accessible magnetic field.

Besides the FMI and AFMI, paramagnetic insulators (PMIs) have also been reported as a source of pure spin currents. SSE signals were observed in $Gd_3Ga_5O_{12}$(GGG)/Pt [22], $DyScO_3$ (DSO)/Pt [22], and $CoCr_2O_4$(CCO)/Pt [23] over the temperature range where GGG, DSO, and CCO are PMIs. In the paramagnetic phase, the concept of magnons is no longer applicable; however, the short-range correlation of spin fluctuations are present [24, 25-26, 27]. Therefore, the observed SSE in PMI/HM heterostructure (as shown in Fig. 1 (b)) must have a different origin, i.e., from spin fluctuations. A complete SSE picture in AFMI must contain ingredients of magnons and spin fluctuations in order to fully account for the data in both AFM and paramagnetic phases.

In this Letter, we report an experimental study of SSE in $FeF_2$/Pt heterostructures. There are two primary reasons why $FeF_2$ is chosen. First, compared with $MnF_2$ which has $H_{SF} \sim 9$ T, $FeF_2$ has stronger uniaxial magnetic anisotropy which gives rise to a larger SF field $H_{SF} \sim 42$ T [28], far greater than the maximum magnetic field used for this study. This ensures negligibly small induced magnetic moment at low temperatures for the magnetic fields applied along the easy axis direction. Second, since the AFM ordering temperature $T_N$ of bulk $FeF_2$ is 78.4 K [29, 30], we can systematically study the SSE response across the antiferromagnetic phase transition. Over the entire temperature range, SSE signals in $FeF_2$/Pt under different magnetic fields up to 12 T show very similar temperature-dependent behaviors. First, there is a SSE peak at ~11.6 K, which can be attributed to the effect of AFM magnons. Second, SSE shows a bump at the $T_N$ of the $FeF_2$ thin films (70 K) and the finite SSE signal extends to 250 K. We can collapse all SSE data above $T_N$ onto a single curve in a scaling plot by using the critical exponent for magnetic susceptibility, providing direct evidence of SSE probing the correlation of spin fluctuations in PMIs.

High quality (110)-oriented $FeF_2$ thin films are grown using molecular beam epitaxy and characterized by reflection high energy electron diffraction and X-ray diffraction (see Supplemental Material Section I). The (110) orientation is chosen so that the easy axis of the AFM spin, i.e., the [001] or the *c*-axis direction, lies in the film plane [31-33]. The uniaxial nature of AFM $FeF_2$ films are characterized by superconducting quantum interference device (SQUID) magnetometry (see Supplemental Material Section II) and by performing field-cooling experiments with the applied magnetic field parallel and perpendicular to $FeF_2$[001] during cooling and measuring the shifted loops of the anisotropic magnetoresistance in FeF$_2$/Co bilayers (see Supplemental Material Section III). The blocking temperature of 50 nm-thick FeF$_2$ thin film is determined to be ~ 70 K from the onset of the exchange bias field and the maximum coercive field in the same field-cooling experiments. The blocking temperature coincides with the temperature where the SSE peak appears in FeF$_2$/Pt; therefore, we believe the blocking temperature is very close to the Néel temperature $T_N$ of the FeF$_2$ films. The deviation of $T_N$ of thin film from the bulk value might be caused by the finite size effect.

To form FeF$_2$/Pt heterostructure for SSE measurements (Fig. 2(a)), 5 nm Pt is directly deposited on top of 50 nm FeF$_2$ with magnetron sputtering and patterned into a Hall bar with dimensions of 100 μm x 630 μm perpendicular to the *c*-axis. Then a 150 nm Al$_2$O$_3$ insulating layer is deposited by electron-beam evaporation, followed by a 50 nm Cr film covering the Hall bar channel area as a heater. In the SSE experiment, a DC current is applied to the Cr heater to generate a vertical temperature gradient across the interface. An external magnetic field is applied in the thin film plane at an azimuthal angle $\phi$ with respect to the AFM easy axis FeF$_2$[001], while the voltage response along the Pt Hall bar channel is recorded as the spin Seebeck signal $V_{SSE}$. Fig. 2(b) plots the $\phi$-dependence of $V_{SSE}$ at 10 K with different heater currents under a 9 T rotating magnetic field. $V_{SSE}$ reaches maximum for the magnetic field along the AFM easy axis (i.e., $\phi = 0°$ and $180°$) and vanishes for the magnetic field perpendicular to the easy axis ($\phi = 90°$). SSE signal shows opposite polarities at $\phi = 0°$ and $180°$, which is consistent with the expectation from the AFM magnon picture. As $\phi$ varies, the magnetic field component projected to the FeF$_2$[001] direction oscillates, so do the Zeeman splitting of the two AFM magnon eigen-modes and the resulting spin current induced SSE signal. In addition, the magnitude of the SSE signal is directly proportional to the heating power (Fig. 2(c)), suggesting the thermoelectric nature of the signal similar to those reported in FMI materials such as YIG [34].

The magnetic field dependence of the SSE signal at 11.6 K is shown in Fig. 2(d). To eliminate parasitic signals, we decompose the $V_{SSE}$ into the symmetric and antisymmetric components: $V_{SSE} = V_{SSE}^S + V_{SSE}^A$, where $V_{SSE}^A = \frac{1}{2}[V_{SSE}(+H) - V_{SSE}(-H)]$ and $V_{SSE}^S = \frac{1}{2}[V_{SSE}(+H) + V_{SSE}(-H)]$, and plot $V_{SSE}^S$ and $V_{SSE}^A$ in Fig. 2(d). $V_{SSE}^S$ could be due to the normal magneto-Seebeck signal produced by an incidental in-plane temperature gradient along the Hall bar channel. Such a background signal is inevitable and also observed in our control sample of MgF$_2$/Pt which does not produce any spin current (see Supplemental Material Section IV). Only $V_{SSE}^A$ is attributed to the SSE signal in the FeF$_2$/Pt heterostructure. When the magnetic field is swept along the easy axis of FeF$_2$, it does not induce any net magnetic moment below < 20 K as the SQUID data indicate; however, it changes the energy splitting between the two AFM magnon eigen-modes. Furthermore, we note that the high-field characteristics of SSE in FeF$_2$/Pt heterostructure are distinctly different from those in YIG/Pt bilayers in the following aspects. First, the SSE voltage in FeF$_2$/Pt increases with the field and shows no sign of saturation up to 12 T. In contrast, the SSE in YIG/Pt decreases with strong magnetic fields due to suppression of thermal magnon population [35]. This effect is not seen in FeF$_2$ since the magnon energy is much higher due to the anisotropy gap. Second, the $V_{SSE}^A$ in FeF$_2$/Pt vanishes at zero magnetic field because the two magnon eigen-modes are degenerate, which leads to zero net spin current; however, the SSE in YIG/Pt remains finite at zero magnetic field due to non-zero population of the sole right-hand magnon mode at finite temperatures [11-15]. We will refer $V_{SSE}^A$ as $V_{SSE}$ hereafter.

The field dependence of the SSE is studied over the temperature range of 3.8 K - 250 K. As the temperature varies, the heater power $P$ also varies due to the heater resistance change, so does the vertical heat current through the sample. To compare the SSE responses at different temperatures, we normalize the SSE voltage signals by the heater power, which is proportional to the spin current density. We plot $V_{SSE}/P$ against both temperature and magnetic field as displayed in Fig. 3(a). $V_{SSE}/P$ curves under different magnetic fields show very similar characteristic temperature dependence as shown in Fig. 3(b). As the temperature increases from 3.8 K, $V_{SSE}/P$ first increases, reaches a maximum at ~11.6 K, and then starts to decrease. Similar low-temperature peak was also observed in MnF$_2$/Pt, but at a lower temperature (~5 K at 8 T), which was later interpreted by Rezende *et al.* [20, 21] in the AFM magnon model. Here we attribute the low-temperature FeF$_2$ peak to the same AFM magnon mechanism. This model qualitatively account for the difference in the peak position between the two AFM films. Since the left-hand magnon branch ($\beta$-mode) lies below the right-hand branch ($\alpha$-mode) under an external field, only the left-hand magnon states are predominately occupied at very low temperatures. As the temperature increases, the left-hand magnon population continues increasing until the right-hand magnon states start to be occupied. Due to the opposite angular momentum of the right-hand magnons, the rising temperature results in a peak in the net spin current and consequently a peak in SSE. In uniaxial AFM materials, the lower the zero-field magnon frequency is, the lower the peak temperature is. In comparison, the zero-field magnon frequencies are 1.6 THz and 0.27 THz in $FeF_2$ and $MnF_2$ respectively [20, 21]; therefore, the AFM magnon peak occurs at a higher temperature in $FeF_2$ than in $MnF_2$. Furthermore, a similar SSE signal is observed in $FeF_2$/Pt for H⊥[001], which has also been reported in the $MnF_2$/Pt bilayer (see experimental details and discussions in Section V of Supplemental Material).

Now we turn to the SSE signal near and above 70 K, i.e., $T_N$ of the $FeF_2$ thin film. As shown in Fig. 3(b), SSE signal shows a second peak at ~70 K, which becomes less sharper as the magnetic field increases but its position remains unchanged. The SSE signal decays above 70 K but remains finite up to 250 K. This is reminiscent of the critical behavior of continuous phase transitions. Since there is no long-range AFM order above $T_N$, the high-temperature SSE signal cannot be interpreted by AFM magnons. Although there is absence of spontaneous magnetic moment above $T_N$, the magnetic field can induce a finite magnetic moment which is proportional to the magnetic susceptibility $\chi$. This can lead to an anomaly in SSE at $T_N$ [36]. Following the critical theory [36], $\chi = t^\gamma f\left(\frac{h}{t^{\beta+\gamma}}\right)$, where $\beta$ and $\gamma$ are the critical exponents for spontaneous staggered magnetization and magnetic susceptibility $\chi$ for $H = 0$ respectively, $h$ is $H/k_B T$, $t = (T - T_N)/T_N$ is the reduced temperature for $T > T_N$, and $f$ is a scaling function. Because of the overlap with the tail of the low-temperature AFM magnon peak below $T_N$, we only analyze the SSE data above $T_N$. In Fig. 3(c), we show that the heating power-normalized SSE data for all magnetic fields above 70 K in a scaling plot: $\frac{V_{SSE}/P}{t^\gamma}$ vs. $\frac{h}{t^{\beta+\gamma}}$ using $\beta = 0.325, \gamma = 1.241$, the critical exponents from the renormalization group calculations for 3D Ising systems [37]. All data collapse onto a single curve in the scaling plot. The excellent agreement indicates that the SSE signal in the PM region measures the magnetic susceptibility which is governed by critical spin fluctuations near $T_N$.

Moreover, we find that the scaling analysis does not work for $\frac{V_{SSE}/P}{t^\beta}$ vs. $\frac{h}{t^{\beta+\gamma}}$ using critical exponent $\beta$ for the staggered magnetization instead of $\gamma$ for the magnetic susceptibility to scale the SSE data with all possible combinations of $\beta$ and $\gamma$ (see Supplemental Material Section VI). Clearly, the SSE signal scales as the magnetic susceptibility $\chi$ rather than field induced sublattice magnetization of the AFMI. Even in a ferro- or ferri-magnet, it is not obvious whether the SSE signal near the Curie temperature scales as the magnetization with critical exponent $\beta$. In fact, SSE measurements in YIG were carried out near the Curie temperature $T_c$ and a $(T_c - T)^3$ power-law behavior was found below $T_c$. This behavior was attributed to the critical behavior of YIG [38]; however, the exponent 3 is not related to the mean-field or critical exponent for the magnetization, nor for the magnetic susceptibility. No satisfactory explanation was given besides acknowledging possible complications due to YIG being a ferrimagnet. Theoretically, the exponent for SSE in ferromagnets was linked to the spontaneous magnetization [39]. Recently, both SSE [22, 23] and spin pumping [40] results in PMI/HM heterostructure were reported, but no quantitative relationship between spin correlation and SSE signal was discussed. Our experimental data and the analyses have established a clear connection between the SSE signal and the magnetic susceptibility near the critical point; therefore, we concluded that SSE is capable of probing correlations of spin fluctuations in magnetic systems.

In summary, we have demonstrated the epitaxial growth of the $FeF_2$(110) thin films on $MgF_2$(110) substrate with controlled AFM anisotropy. The exchange bias data confirm that the uniaxial AFM spin axis is along the $FeF_2$[001] direction and the AFM ordering temperature is 70 K. The temperature dependence of the SSE signal shows two peaks located at 11.6 K and 70 K respectively. The former is attributed to the AFM magnons, and the latter to the enhanced correlation of critical fluctuations near the AFM ordering temperature. Our results suggest that both magnons in magnetically ordered phases and the correlated spin fluctuations near phase transitions can act as pure spin current sources. This picture provides an alternative interpretation of the enhanced SSE and spin pumping signals near the AFM ordering temperature previously observed in FMI/AFM/NM heterostructures [5, 6, 41].

We acknowledge the useful discussions with Ran Cheng, Mark Lohmann, and Wei Yuan. This work was supported as part of the SHINES, an Energy Frontier Research Center funded by the US Department of Energy, Office of Science, Basic Energy Sciences under Award No. SC0012670. This work was also supported by the Startup Funds from University of California, Riverside.

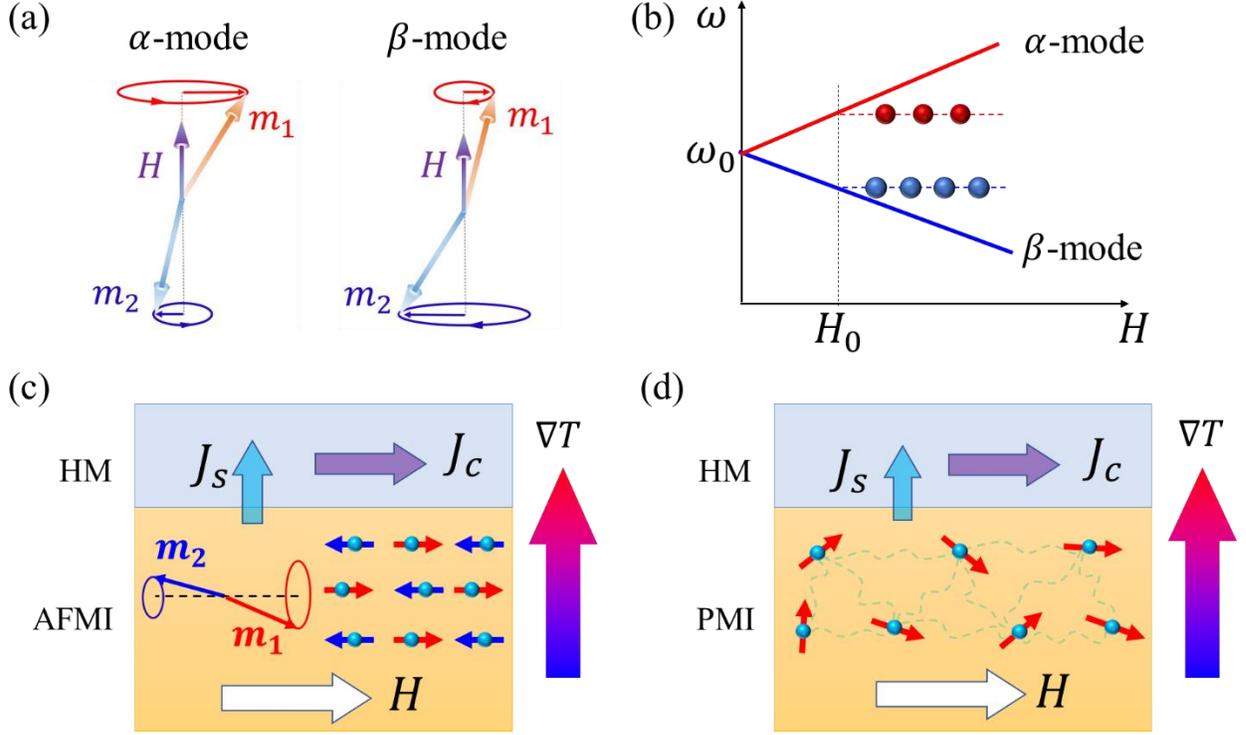

FIG. 1. (a) Illustration of $\alpha$- and $\beta$-modes of magnons in an uniaxial AFM. $m_1$ and $m_2$ are magnetic moments on two spin sublattices, respectively. (b) Frequency $\omega$ of AFM magnons for the $\alpha$- and $\beta$-modes as a function of the magnetic field $H$ applied along the AFM easy axis. The red and blue symbols indicate the occupation of $\alpha$ and $\beta$ AFM magnons, respectively. Schematic diagrams of SSE measurements for AFMI/HM (c) and PM insulator/HM (d) heterostructures. $\nabla T$ is the temperature gradient; $J_s$ is the spin current density across the interface and $J_c$ is the charge current density generated by the inverse spin Hall effect.

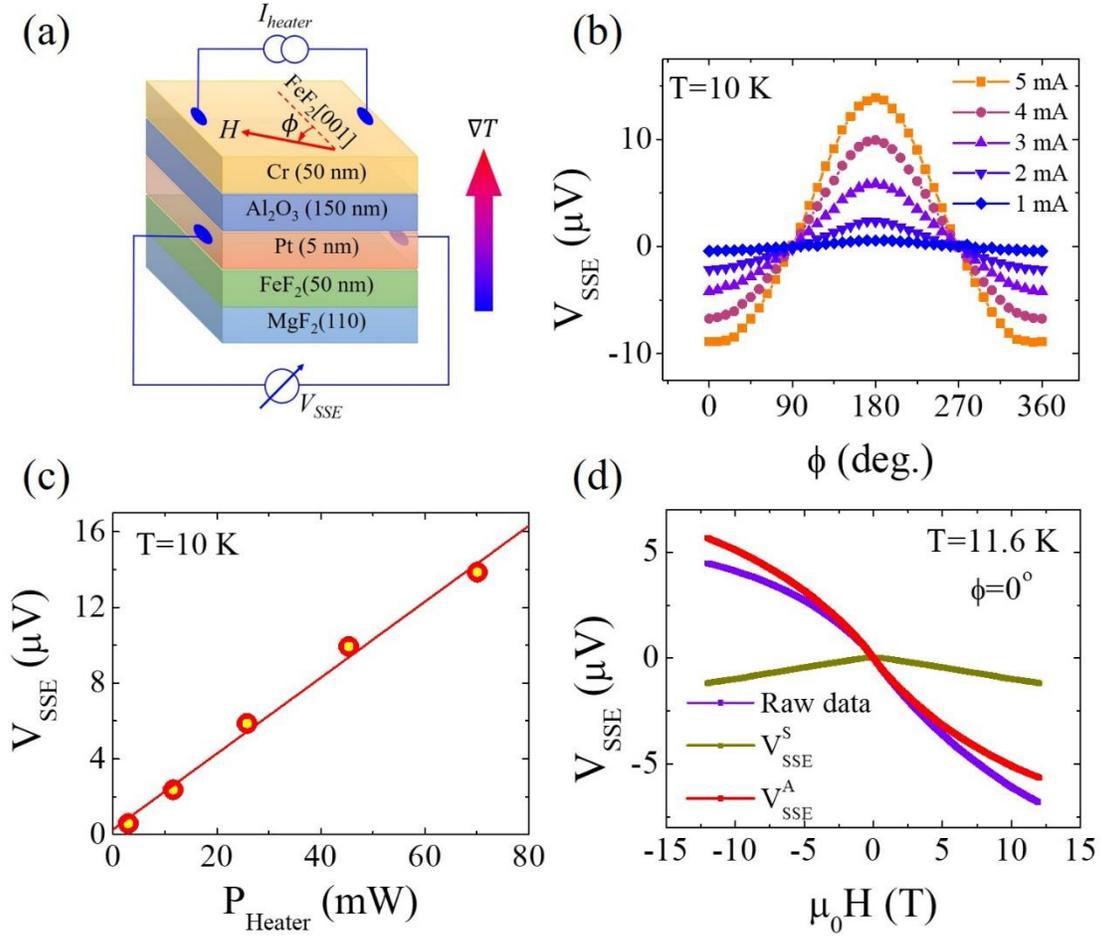

FIG. 2. SSE signal $V_{SSE}$ in FeF$_2$(50 nm)/Pt(5 nm) heterostructure. (a) Schematic diagram of the sample structure and longitudinal SSE measurement geometry. A current in Cr generates a temperature gradient $\nabla T$ which produces $V_{SSE}$ in Pt. $H$ is applied at an angle $\phi$ to the FeF$_2$[001] direction in the film plane. The patterned Pt strip is 100 μm wide and 630 μm long and the $V_{SSE}$ is measured along the length of the strip. (b) $\phi$-dependence of $V_{SSE}$ at 10 K for different heater currents with a constant magnetic field $H$= 9 T. (c) $V_{SSE}$ vs. heating power $P$ at 10 K. (d) $V_{SSE}$ as a function of $H$ applied along FeF$_2$[001] at $T$=11.6 K with a heater current of 3 mA. Purple, red, and brown curves represent the raw data, antisymmetric, and symmetric components, respectively.

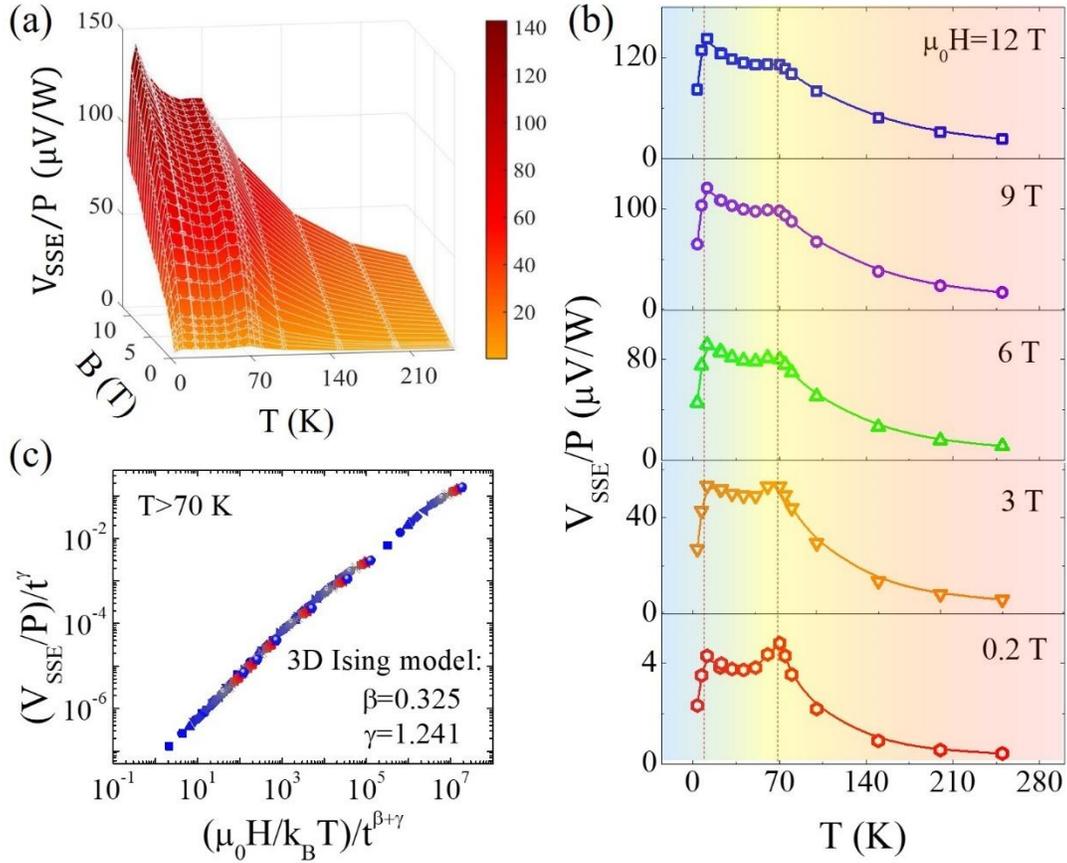

FIG. 3. Temperature dependence of $V_{SSE}$ in FeF$_2$/Pt heterostructure. (a) Three-dimensional plot of $V_{SSE}/P$ (only the antisymmetric component) as a function of $H$ and $T$. (b) Temperature dependence of $V_{SSE}/P$ for $H=$ 12, 9, 6, 3 and 0.2 T from top to bottom indicated by different colors. Two vertical dashed lines indicate the peak positions. Solid curves are guides to the eye. The lowest temperature in our experiments is 3.8 K. (c) Scaling plot of all the SSE data above $T_N$ (70 K). Different colors represent data for different magnetic fields as shown in (b). We adopt $\beta = 0.325$ and $\gamma = 1.241$, which are the renormalization group critical exponents for three-dimension Ising systems.